\documentstyle[epsfig,11pt]{article}
\title{\bf Hadron-nucleus scattering in the local reggeon model with
pomeron loops for realistic nuclei}
\author{M.A.Braun, A.Tarasov  \\
S.Peterburg State University, Russia}
\begin{document}
\maketitle
\input epsf
\def\beq{\begin{equation}}
\def\eeq{\end{equation}}
\def\phid{\phi^{\dagger}}
\def\pd{\partial}
\def\by{\bar{y}}
\def\tt{\tilde{T}}

{\bf Abstract}

Contribution of simplest loops for hadron-nucleus scattering
cross-sections is studied in the Local Reggeon
Field Theory  with a supercritical pomeron.
It is shown that inside the nucleus  the supercritical
pomeron transforms into a subcritical one, so that perturbative
treatment  becomes possible. The pomeron intercept becomes complex,
which leads to oscillations in the cross-sections.

\section{Introduction}
High-energy hadron-nucleus scattering has long been studied
in the framework of the local Regge-Gribov theory with a
self-interacting supercritical pomeron. The first successful
description of the total cross-sectuion in this approach was
presented by A.Schwimmer in 1975 ~\cite{schwim} who performed
summation of fan diagrams in the limit when the pomeron slope
$\alpha'$ was zero. With the advent of the QCD
attention has been shifted to self-interacting non-local BFKL
pomerons. In this approach an analog of the Schwimmer model
has been constructed in the form of the
Balitsky-Kovchegov evolution equation ~\cite{bal,kov,bra1}.
However QCD can reliably describe only the hard region of the
dynamics. Soft processes, contributing to the bulk of the total
cross-section, are better described by the old-fashioned
local reggeon theory, so that the latter remains a useful tool
in spite of its venerable age.

Both the old local and new non-local pomeron models have been fully
considered and applied only in the quasi-classical tree approximation. Loops have
been  neglected in both formalisms.
This  approximation can be
justified if the parameter $\gamma=\lambda\exp{\Delta y}$ is small,
with $y$ the rapidity and $\Delta$ and $\lambda$ the pomeron intercept
and triple pomeron coupling.
Then for a large nuclear target, such that $A^{1/3}\gamma\sim 1$,
the tree diagrams indeed give the dominant contribution and loops
may be dropped. However with the growth of $y$ the loop
contribution becomes not small and this approximation breaks down.

Full calculation of the loop contribution seems to
be a formidable task  for the non-local QCD pomeron.
So it seems to be worthwhile to start with the Local
Reggeon Field Theory (LRFT) with a
supercritical pomeron. Such a study, apart from its possible lessons for the
modern QCD approach, also has an independent value, since, as mentioned, the
old LRFT with phenomenological parameters describes the
soft dynamics of high-energy strong interactions not so badly.
Much work has been done on
the influence of loop diagrams in the zero-dimensional LRFT,
both   theoretically old ago in ~\cite{amati,aless, jengo, ciabel} and
numerically recently in ~\cite{bravac}. This influence turns out to
be decisive for the asymptotic behaviour at large energies and transforms the
supercritical  pomeron into a weakly subcritical one with the effective
intercept $\propto -\exp(1/\lambda^2)$ where $\lambda$ is the
triple-pomeron coupling  assumed to be small.

Unfortunately generalization of these beautiful results to the realistic
case of two transverse dimensions is prohibitively difficult. To start
with, one is forced to introduce a non-zero value of the slope: otherwise the
loop contribution is divergent in the impact parameter.  However even in the
tree approximation solution of the model with $\alpha'\neq 0$ is only
possible  numerically. Second, the model in $d_T=2$ needs renormalization
in the ultraviolet. And, most important, the method used to solve the
model in $d_T=0$, which is to study the corresponding quantum-mechanical
system   and the relevant Schroedinger equation, is inapplicable in the
realistic case, since instead of ordinary differential equations one arrives
at  equations with functional derivatives. In fact  summation
of all loop contributions is equivalent to  a complete solution of the
corresponding  quantum field theory, the task which seems to be beyond
our present possibilities. So at most one can hope to obtain some partial
results which might shed light onto the properties of the model with loops.
There were several attempts to study the high-energy behaviour of the
LRFT with a supercritical pomeron using different approximate
techniques and giving contradicting results ~\cite{abarb}, ~\cite{ciaf}.

In our previous study, instead of trying to solve
the model for the purely hadronic scattering we considered
the hadron-nucleus scattering and propagation of the pomeron inside
the heavy nucleus target ~\cite{BT}. Moreover to avoid using numerical
solution of the tree diagrams contribution with
diffusion in the impact parameter, we concentrated on the case of
a constant nuclear density which allowed to start with the known
analytical solutions. We have found that the
nuclear surrounding  transforms the pomeron from the supercritical
one with intercept $\epsilon>0$ to a subcritical one with the intercept
$-\epsilon$. Then Regge cuts, corresponding to loop diagrams, start at
branch points located to the left of the pomeron pole and their contribution
is subdominant at high energies. As a result the theory aquired the
properties similar to the standard LRFT with a subcritical pomeron and allows for
application of the perturbation theory.

However the adopted approximation of a constant nuclear density in
the whole space was obviously too crude. In fact it did not allow to make
convincing conclusions about the physical cross-sections, so that we
had to recur to some weakly supported guesses on this point.
In this paper we try to improve our treatment and consider realistic
nuclear targets which occupy finite space volume. This enables us to
predict the influence of loops on the total cross-section and study
applicability of the perturbative approach to hadron-nucleus scattering
with pomeron loops.

Our results depend on the behaviour of the nuclear profile
function $T(b)$ at large values of the impact parameter $b$.
For a finite nucleus with $T(b)=0$ at $b>R_A$ all our conclusions
found in ~\cite{BT} for constant $T(b)$ remain valid and one can apply
perturbative methods to study the loop contribution. However for an
infinite nucleus, for which $T(b)$ exponentially falls with $b$ but never
vanishes at finite $b$, the loop influence is perturbative only inside
the nucleus, at not too large $b$. At very large $b$, well outside the bulk
of the nuclear matter, the loop contribution blows up and the cross-section
becomes unperturbative.

The paper is organized as follows. In the next section we introduce
the model and reformulate it to describe the loops in the nuclear
surrounding. In Section 3 we introduce a quasi-local approximation
to constructively treat the case of realistic nuclei.  In Section 4
we formulate the appropriate Dyson equation for the amplitude with loops.
Our numerical results for total pPb and pCu cross-sections are presented
in Section 5. Finally Section 6 draws some conclusions.

\section{Local pomeron in the nuclear field}
The LRFT model is based on two pomeron fields
$\phi(y,b)$ and $\phid(y,b)$ depending on the rapidity $y$ and
impact parameter $b$, with a Lagrangian density
 \beq
L=L_0+\lambda\phid\phi(\phi+\phid)+g\rho\phi.
\label{lagden}
\eeq
Here the free Lagrangian density is
\beq
L_0=\phid\Big(\frac{1}{2}\stackrel{\leftrightarrow}{\partial_y}
-\alpha'\nabla_b^2+\epsilon\Big)\phi\equiv \phid S\phi
\label{freel}
\eeq
where $\epsilon$ is the intercept minus unity and $\alpha'$ is the slope.
The  source term describing interaction with the nuclear target at low
energies is
\beq
g\rho(y,b)=gAT(b)\delta(y).
\eeq
where $g$ is the pomeron-nucleon coupling constant and $T(b)$ the
profile function of the nucleus.
For a supercritical pomeron $\epsilon>0$ and $\lambda<0$.

The classical equation of motion are
\beq
\frac{\delta
L}{\delta\phi}=-\partial_y\phid-\alpha'\nabla_b^2\phid
+\epsilon\phid+\lambda{\phid}^2+2\lambda\phi\phid +g\rho=0
\label{eqphidi}
\eeq
and
\beq
\frac{\delta
L}{\delta\phid}=\partial_y\phid-\alpha'\nabla_b^2\phi
+\epsilon\phi+\lambda\phi^2+2\lambda \phid\phi=0.
\label{eqphi}
\eeq
From the
latter equation we find $\phi=0$ and the equation for $\phid$
takes the form
\beq
\partial_y\phid=-\alpha'\nabla_b^2\phid
+\epsilon\phid+\lambda{\phid}^2,
\label{eqphid}
\eeq
with an initial condition
\beq
\phid(y=0)=g\rho
\label{iniphid}
\eeq
Equation (\ref{eqphid}) describes evolution of the pomeron field in
rapidity and its diffusion in the impact parameter inside the nucleus.
We denote the solution of the classical equation of motion
(\ref{eqphid}) with the initial condition (\ref{iniphid}) as
$\xi(y,b)$

To go beyond the classical approximation and thus study loops we
make a shift in the quantum field $\phid$:
\beq
\phid(y,b)=\phid_1(y,b)+\xi(y,b)
\label{shift}
\eeq
and reinterprete our theory in terms of fields $\phi$ and $\phid_1$.
In the  Lagrangian terms linear in $\phi$
vanish due to the equation of motion for $\xi$ and
we obtain
\beq
L=\phid_1(S+2\lambda \xi)\phi+\lambda\xi\phi^2+
\lambda\phid_1\phi(\phid_1+\phi).
\eeq
This Lagrangian corresponds to a theory in the vacuum with the
pomeron propagator  in the external field $f(b,y)=2\lambda\xi(y,b)$
\beq
P =-(S+2\lambda\xi)^{-1},
\label{prop}
\eeq
the standard triple interaction
and extra interaction described by the term $\lambda\xi\phi^2$. This new
interaction corresponds to  transition of a pair of pomerons into the vacuum
at point $(y,b)$ with a vertex $\lambda\xi(y,b)$, see Fig. 1.

\begin{figure}
\hspace*{4 cm}
\epsfig{file=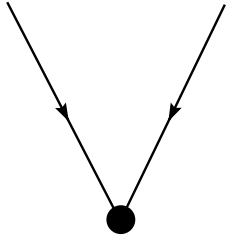,width=3 cm}
\caption{The new vertex for two-pomeron annihilation, which appears
after the shift in field $\phid$}
\label{fig1}
\end{figure}

Loops  can be formed both by the standard interaction
and the new one. In the latter case they are to be accompanied by at
least a pair of of standard interactions.
Diagrams with a few  simple loops in the Green function
are illustrated in Fig. 2.
One immediately observes that a loop formed by the standard
interaction has the order $\lambda^2/\alpha'$ and requires renormalization.
A loop formed by the new interaction has the order $\lambda^3/\alpha'$ and
is finite.

\begin{figure}
\epsfig{file=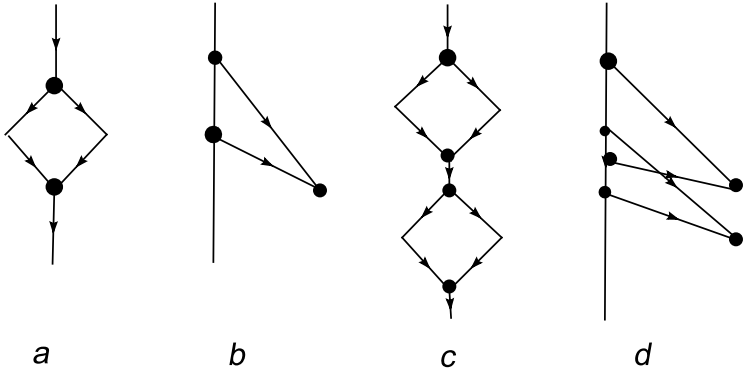,width=12 cm}
\caption{Some simple loop diagrams for the pomeron Green function}
\label{fig2}
\end{figure}

The amplitude is obtained as a tadpole $g<\phid_1(y,b)>$. The simplest
diagrams for it contain one loop and are shown in Fig. 3 $a,b$. Diagrams
with more loops are shown in Figs. 3 $c,d$.

\begin{figure}
\hspace*{2 cm}
\epsfig{file=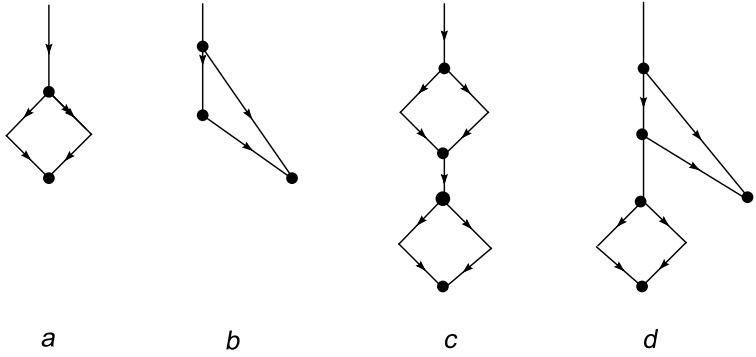,width=10 cm}
\caption{Diagrams with one loop ($a,\, b$) and two loops ($c$ and $d$) for the
scattering amplitude}
\label{fig3}
\end{figure}

Propagator  (\ref{prop}) in the external field
$f(y,b)=2\lambda\xi(y,b)$ where $\xi(y,b)$ is the classical
field $\phid$ satisfies the equation
\beq
\frac{\partial P(y,b|y'b')}{\partial y}=
(\epsilon-\alpha'\nabla_b^2) P(y,b|y',b')+
f(y,b)P(y,b|y',b')
\label{eqprop}
\eeq
with the boundary conditions
\beq
P(y,b|y',b')=0,\ \ y-y'<0,\ \ P(y',b|y',b')=\delta^2(b-b').
\label{inigfb}
\eeq
In the general case  propagator $P$ can only be calculated
numerically, just as the external field $f(y,b)$. Its analytic form
can be found in two cases ~\cite{BT}.

If the slope $\alpha'=0$ then one finds
\beq
f(y,b)=
-\frac{\epsilon a(b)e^{\epsilon y}}{p(y,b)}
\label{fyb}
\eeq
and
\beq
P(y,b|y',b')=\delta^2(b-b')P_0(y,y',b),\ \
\ \ P_0(y,y',b)=e^{\epsilon(y-y')}\frac{p^2(y',b)}{p^2(y,b)},
\label{prop0}
\eeq
where
\beq
a(b)=-\frac{\lambda g\rho(b)}{\epsilon} >0
\label{adef}
\eeq
and
\beq
p(y,b)=1+a(b)\Big(e^{\epsilon y}-1\Big).
\label{pdef}
\eeq
Propagator $P_0(y,y',b)$ corresponds to the zero-dimensional theory
with $a(b)$ as an external parameter.
It is
remarkable that at large $y$ propagator
$P_0(y,y',b)\simeq \exp(-\epsilon y)$, that is behaves as
the free propagator with the opposite sign of $\epsilon$ and
so vanishes at $y\to\infty$. In contrast to the free propagator
it corresponds to a subcritical pomeron.

The second case which admits an analytic solution for the propagator is
that of the nuclear matter, that is the case when the profile function is
constant in all the transverse space
\beq
T(b)=T_0.
\eeq
Then the classical field $\xi$ and the external field $f=2\lambda\xi$
become $b$-independent and the propagator in this
field takes the form
\beq
P(y,b|y',b')=\frac{1}{4\pi\alpha' (y-y')}
e^{\epsilon (y-y')-\frac{(b-b')^2}{4\alpha'(y-y')}}
\frac{p^2(y')}{p^2(y)}.
\label{prop1}
\eeq
where $p(y)$ is given by (\ref{pdef}) with a constant $a$.
Note an especially simple case when $a=1$ and $p(y)=\exp{\epsilon y}$.
Then the propagator
in the external  field coincides with the free propagator with the
opposite sign of $\epsilon$, so that
the theory formally corresponds to a subcritical pomeron model
with an additional interaction shown in Fig. \ref{fig1}

\section{Realistic nucleus}
The full treatment of the scattering on a realistic nucleus with
the profile function $T(b)$ which varies with $b$ and vanishes as
$b\to \infty$ requires solution of the equations (\ref{eqphid}) for
the pomeron field and (\ref{eqprop}) for the propagator in the
found external field $f(y,b)$. Both tasks can be done only
numerically and look hardly feasible.

To facilitate the problem we shall use the approximation of a weakly
varying $T(b)$ . Equivalently we assume that the slope $\alpha'$,
which governs diffusion in the impact parameter, is small.
Accordingly we shall put $\alpha'$ to zero wherever possible.
In particular we shall use expression (\ref{fyb})
for the field at fixed value $b$
of the impact parameter. For the  propagator we
shall use expression (\ref{prop0}) so long as it
stands alone. However we cannot do this in pomeron loops, where it is
squared leading to a divergence  at $\alpha'\to 0$.

For loops we shall restrict
ourselves with the leading term as $\alpha'\to 0$
and so use the leading approximation for the
propagator inside the loop in this limit but not the limit itself.
This leading approximation obviously coincides with the
approximation of a slowly varying nuclear field and is given by
Eq. (\ref{prop1}) with $p$ depending on $b$ in accordance with
definition (\ref{pdef}).
One can see it from the following reasoning. Let us seek the
propagator satisfying Eq. (\ref{eqprop}) as a product
\beq
P(y,b|y',b')=P_0(y,y',b)P_1(y,b|y'b')
\eeq
with the known $P_0(y,y',b)$ defined in Eq. (\ref{prop0}).
Then we find an equation for $P_1$
\[P_0(y,y',b)\frac{dP_1(y,b|y',b')}{dy}=
\alpha'\nabla_b^2 P_0(y,y',b)P_1(y,b|y',b).\]
Assuming the smallness of $\alpha'$ and non-singularity of
$P_0$ as a function of $b$ or equivalently a weak dependence
of $P_0$ on $b$ we can neglect action of the derivatives in $b$
on $P_0(y,y',b)$ to obtain an equation for $P_1$
\[\frac{\partial P_1(y,b|y',b')}{\partial y}=
\alpha'\nabla_b^2 P_1(y,b|y',b).\]
Solution of this equation brings us to propagator (\ref{prop1}).

With this approximation we now calculate the simplest loops which
appear in the theory paying special attention to their $b$-dependence.

The simple loop Fig. \ref{fig2}$a$ is given by the expression
\beq
\Sigma_1(y,b|y',b')=-2\lambda^2P^2(y,b|y',b')+\Delta\epsilon\delta^2(b-b')
\delta(y-y').
\label{sigma11}
\eeq
The second term comes from the intercept renormalization. Using
(\ref{prop1}) we find
\beq
\Sigma_1(y,b|y',b')=-2\lambda^2\frac{p^4(y',b}{p^4(y,b)}
\Big(\frac{1}{4\pi\alpha' (y-y')}\Big)^2
e^{2\epsilon (y-y')-\frac{(b-b')^2}{2\alpha'(y-y')}}
+\Delta\epsilon\delta^2(b-b')\delta(y-y').
\label{sigma12}
\eeq
The exponential factor depending on $(b-b')^2$ obviously becomes
proportional to $\delta^2(b-b')$ in the limit $\alpha'\to 0$
\beq
\frac{1}{2\pi\alpha'(y)}
{e^{-\frac{b^2}{2\alpha'y}}}_{\alpha'\to 0}\to\delta^2(b).
\eeq
Thus we obtain at small $\alpha'$
\beq
\Sigma_1(y,b|y',b')=-\delta^2(b-b')\frac{\lambda^2}{4\pi\alpha' (y-y')}
e^{2\epsilon (y-y')}\frac{p^4(y',b}{p^4(y,b)}
+\Delta\epsilon\delta^2(b-b')\delta(y-y').
\label{sigma13}
\eeq
Note that at fixed $b$ and rapidities $\Sigma_1$ is finite.
Ultraviolet divergence appears in the process of integration
over rapidity at $y-y'=0$. Regularizing this integration by
cutting it at $y-y'=y_{min}$ one can express
\beq
\Delta\epsilon=-\frac{\lambda^2}{4\pi\alpha'}\ln (c_Ry_{min})
\eeq
with $c_R$ a renormalization constant taking  place
of $\Delta\epsilon$.

Now we pass to the simplest new self-mass of Fig. \ref{fig2}$b$.
It does not need renormalization and is given by
\beq
\Sigma_2(y,b|y',b')=4\lambda^3P(y,b|y'b')\int dzd^2c\,\xi(z,c)
P(y,b|z,c)P(y',b'|z,c).
\label{sigma21}
\eeq
With the approximate form (\ref{prop1}) for the propagator we find
a product of exponential factors depending on impact parameters
\[\frac{1}{(4\pi\alpha')^3(y-y')(y-z)(y'-z)}
e^{-\frac{(b-b')^2}{4\alpha'(y-y')}-\frac{(b-c)^2}{4\alpha'(y-z)}
-\frac{(b'-c)^2}{4\alpha'(y'-z)}}\]
to be integrated over $z$ and $c$. Integration over $c$ converts
this into
\[\frac{1}{(4\pi\alpha')^2(y-y')(y+y'-2z)}
e^{-\frac{(b-b')^2}{4\alpha'(y-y')}-\frac{(b-b')^2}{4\alpha'(y+y'-2z)}}
\to \frac{1}{8\pi\alpha'}\delta^2(b-b')\ \ {\rm at}\ \ \alpha'\to 0.\]
So in our approximation we find the new self-mass
\beq
\Sigma_2(y,b|y',b')=-\delta^2(b-b')\theta(y-y')
\frac{4\lambda^2\epsilon a(b)}
{8\pi\alpha'}
\int_0^y\frac{dz}{y-z}e^{\epsilon(2y-z)}\frac{p^3(z,b)}{p^4(y,b)}.
\label{sigma22}
\eeq

\section{Amplitude at fixed impact parameter}
In the adopted approximation all  integrations over intermediate
impact parameters are performed with the help of $\delta$ functions
and at fixed impact parameter $b$ the final amplitude $T(y,b)$ is a function
of $b$ presented as a mutiple integral over intermediate rapidities.
So the whole picture is local in $b$.

The propagator (\ref{prop0}) in the external field has the crucial
property that it decreases with $y$ as $y\to\infty$:
\beq
P(y,y',b)_{(y-y')\to\infty}\to e^{-\epsilon(y-y')}f(y',b).
\eeq
Because of this property internal integrations over rapidities do
not lead to contributions which grow faster than the lowest order term.
This ensures convergence of the perturbative expansion at small enough
values of $\lambda$

Introduction of loops into the amplitude ${\cal A}$ is achieved by the
Dyson equation
\beq
{\cal A}(y,b)={\cal A}^{(0)}(y,b)+
\int_0^y dy_1\int_0^{y_1}dy_2P(y,y_1,b)\Sigma(y_1,y_2,b){\cal A}(y_2,b),
\label{trand}
\eeq
where the lowest order term is
\beq
{\cal A}^{(0)}(y,b)=-\frac{a(b)e^{\epsilon y}}{\lambda p(y,b)}
\label{amp0}
\eeq
with $p(y,b)$ given by (\ref{pdef})
and
the pomeron self mass $\Sigma(y,b|y'b')$ presented in our approximation
as
\beq
\Sigma(y,b|y',b')=\delta^2(b-b')\Sigma(y,y',b)
\eeq

In accordance with the convergence of the perturbative
expansion,  in the first approximation  we take
\beq
\Sigma(y,b|y',b')=\Sigma_1(y,b|y',b')+\Sigma_2(y,b|y',b')
\eeq
where $\Sigma_1$ and $\Sigma_2$ are the 2nd and 3d order
self-mass contributions studied above.
We present
\beq
{\cal A}(y,b)={\cal A}^{(0)}(y,b)r(y,b).
\eeq
The equation for $r(y)$ takes the form
\beq
r(y,b)=1+X_1(y,b)+X_2(y,b),
\label{eqr}
\eeq
where $X_1$ and $X_2$ are parts coming from $\Sigma_1$ and
$\Sigma_2$ respectively.
Explicitly
\beq
X_1=-\frac{C}{p(y,b)}\ln(c_Ry)
\int_0^ydy_1p(y_1,b)r(y_1)-\frac{C}{p(y,b)}\int_0^y\frac{dz}{z}
\Big(\omega(y,z,b)-\omega(y,0,b)\Big),
\label{amprand1}
\eeq
where
\beq
C=\frac{\lambda^2}{4\pi\alpha'}
\eeq
and
\beq
\omega(y,z,b)=e^{\epsilon z}\int_z^ydy_1
\frac{p^3(y_1-z,b)}{p^2(y_1,b)}r(y_1-z,b).
\label{lomega}
\eeq

The part $X_2(y)$ is
\beq
X_2(y)=-\frac{2a(b)\epsilon C}{p(y,b)}\int_0^ydy_1
\frac{e^{\epsilon y_1}}{p^2(y_1,b)}
\int_0^{y_1}\frac{dy_3}{y_1-y_3}p^3(y_3,b)e^{-\epsilon y_3}
\int_{y_3}^{y_1}dy_2\frac{e^{\epsilon y_2}}{p(y_2,b)}r(y_2,b).
\label{amprand2}
\eeq

Eq. (\ref{eqr}) contains the renormalization constant $c_R$ as
a parameter. To fix it we use the prescription proposed in ~\cite{BT}.
The contribution to ${\cal A}(y,b)$ of a single loop is
\beq
{\cal A}^{(1)}(y,b)={\cal A}^{(0)}(y,b)r^{(1)}(y,b)
\eeq
where
\beq
r^{(1)}(y,b)=\Big(X_1(y,b)+X_2(y,b)\Big)_{r(y,b)=1}.
\label{r1}
\eeq
The right-hand side of Eq. (\ref{r1}) can be  calculated
and shown to have a finite limit at $y\to\infty$
\beq
r^{(1)}(y,b)_{y\to\infty}=-
\frac{\lambda^2}{4\pi \alpha'\epsilon}
\Big(\ln\frac{c_R}{2\epsilon}+1-C_E\Big),
\label{limr1}
\eeq
where $C_E$ is the Eiler constant. Remarkably this limit does not depend
on $b$ and is universal. So one can use it to fix the
renormalization constant by requirement that this limit is zero:
\beq
c_R=2\epsilon e^{C_E-1}.
\eeq
This means that at large rapidities the loop contribution vanishes,
so that one defines the pomeron intercept from the
asymptotic of the amplitude requiring it to coincide with the
lowest order term.

With all this, one should be careful in applying these results for different
values of impact parameter $b$. The decrease of the propagator $P(y,y',b)$
obviously starts when $p(y,b)$ begins to rise, which requires
$ a(b)e^{\epsilon y}>>1$ or
\[ e^{\epsilon y}>>\frac{1}{a(b)}\]
It follows that for very small $a(b)$ the actual decrease will start at
very large rapidities. At \\ $a(b)e^{\epsilon y}\sim 1$ the propagator
will have order $1/a(b)$.  In other words, at small $a(b)$
we expect the ratio $r(y,b)$ to rise with rapidity from unity to a large
value $1/a(b)$ at $\epsilon y=y_0\sim -\ln(a(b))$ and then drop to zero
at $y>>y_0$. At its maximum the loop contribution will be greater
than the lowest order by factor
\beq
\zeta=\frac{\lambda^2}{4\pi\alpha'\epsilon a(b)},
\label{zeta}
\eeq
where $\epsilon$ in the denominator comes from the essential
interval of integration over intermediate rapidities.
If the triple pomeron coupling constnt $\lambda$ is
small enough this does not prevent using perturbation theory since
$\lambda^2/a(b)\sim \lambda$. However it does bring difficulties if
one fixes $\lambda$ and then varies $a(b)$ to very small values at
large $b$. This point is essential in the application of the model
to realistic nuclei studied in the next sections.

\section{Numerical results}
The total hadron-nucleus
cross-section is given by  twice
the imaginary part of the hA amplitude, integrated over all values
of the impact parameter. To take into account unitarity at fixed $b$
we glauberize the contribition from a single fan with loops,
which implies that the projectile hadron may emit any number of such
structures.
\beq
\sigma(y)=2\int d^2b\Big(1-e^{g{\cal A}(y,b)}\Big).
\label{sigma}
\eeq

To calculate ${\cal A}(y,b)$ we have solved the Dyson equation (\ref{trand})
for proton-lead and proton-copper collisions ($A=207$ and 64) at each
value of $b$
with the pomeron self-mass given by the sum of lowest order loops
$\Sigma_1+\Sigma_2$.

We used more or less standard values for the pomeron parameters extracted
from the experimental data on proton-proton scattering
\beq
\epsilon=0.08, \ \ \alpha'=0.2\ \ GeV^{-2},\ \ \lambda=-0.48
GeV^{-1},\ \ g=5.94\ \ GeV^{-1}.
\label{par}
\eeq
Note that $\epsilon$ and $g$ can be rather reliably
taken directly from the proton-proton inelastic cross seections.
The rest of the parameters is not so well established and model
dependent. Our set agrees
with the parametrization used in ~\cite{armesto}.
With the set (\ref{par}) the value of parameter $\zeta$
which controls the relative order of loop contribution and so applicability
of perturbative treatment is of the order $1/a(b)$. For heavy nuclei
the maximal value of $a(b)$  is $a(0)\sim 1$. So strictly speaking
for realistic nuclear scattering the used values of $\lambda$
do not lie within the applicability of perturbative treatment.
This circumstance has to be taken into account in analyzing our results
with $\lambda$ given in (\ref{par}), which should be taken as qualitative.

\begin{figure}
\hspace*{2 cm}
\epsfig{file=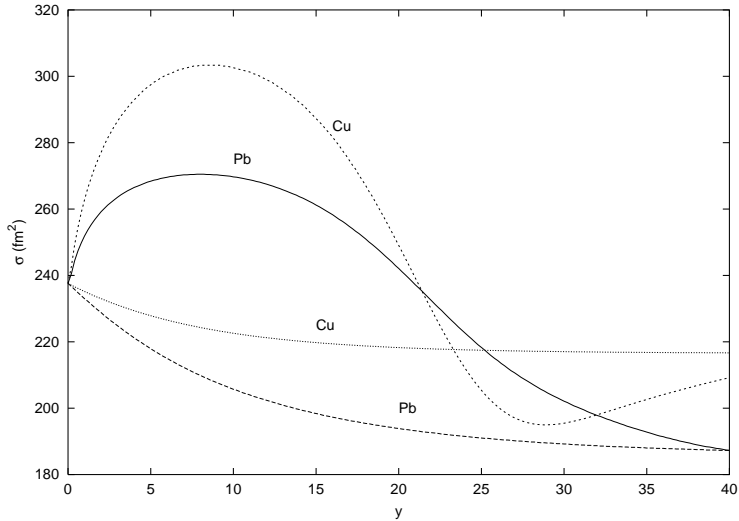,width=10 cm}
\caption{pPb and pCu cross-sections on the target nucleus as a sphere
with a constant density. The oscillating (monotonous) curves corresponds
to presence (absence) of pomeron loops. The pCu cross-sections
are multiplied by 2.55 to coincide with the pPb one at small $y$ }
\label{fig4}
\end{figure}

We considered two models for the spatial structure of the nuclear target.
The simplest one assumes the nucleus to be a sphere with a constant
nuclear density of radius
\[ R_A=A^{1/3}R_0,\ \ r_0=1.15 \ \ {\rm fm}.\]
This is the nuclear model most suitable for our approach, since it
excludes the region of very small nuclear densities ouside the nucleus
in which the perturbative treatment is impossible due to large values
of $\zeta$.
In this case the found hA cross-sections are shown in Fig. \ref{fig4}
as a function of rapidity $y$. To see the influence of loops we also
presented the cross-sections without loop contribution, which correspond to
the monotonous curves. To clearly see the $A$-dependence we rescaled the
pCu cross-section to be equal to the pPb one at small $y$.
As one observes, the loop contribution is quite
noticeable and leads to oscillations in the cross-sections. The origin
of oscillations is understandable from the fact that with
loops the original pomeron pole splits into two complex
conjugated poles, which remain on the physical sheet due to the
absorptive nature of the loop
(negative signs in (\ref{sigma13}) and (\ref{sigma22})).
The $A$ dependence is quite strong and also understandable.
With the growth of $A$ the damping effect of the nuclear surrounding grows,
so that the cross-sections generally become relatively smaller, both
with and without loops.

\begin{figure}
\hspace*{2 cm}
\epsfig{file=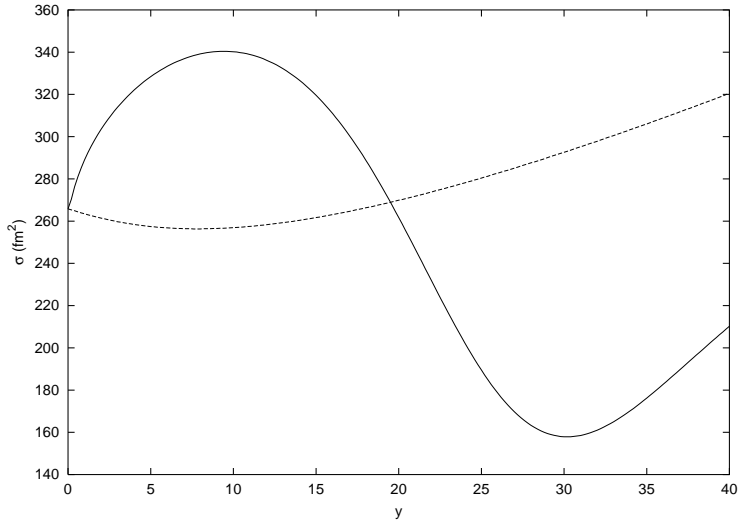,width=10 cm}
\caption{pPb cross-sections on the target nucleus with the Woods-Saxon
density. The oscillating (monotonous) curve corresponds
to presence (absence) of pomeron loops }
\label{fig5}
\end{figure}

As a second model for the nucleus density we considered the standard
Woods-Saxon one. Such a nucleus has no finite dimension and we have to
vary $b$ from zero to infinity. With the growth of $b$ outside the nucleus
parameter $a(b)$ becomes exponentially small and formal application of
our formulas becomes impossible because of the growth of the perturbative
parameter $\zeta$. To overcome this difficulty we  take into account
that at $b$ outside the nucleus the projectile practically interacts with
a free nucleon and so its aplitude is wholly determined by the single
pomeron exchange without any loop insertions, which are assumed to be
incorporated into the effective pomeron parameters. So we have introduced
a cutoff into the loop contributions multiplying them by
\[ c_f=\frac{a(b)}{a_0+a(b)} \]
and choosing $a_0=a(R_A+\Delta b)$ with $b=1.5$ fm.
The resulting pPb cross-sections are presented in Fig. \ref{fig5} again
together with the cross-sections without loops, which correspond to the
monotonous curve. The found cross-sections
show the same qualitative behaviour as with the finite nucleus, although the
absolute values and the amplitude of oscillations are somewhat greater.
The cross-sections without loops naturally grow faster that
for a finite nucleus because of the contribution of the nuclear halo.

\begin{figure}
\hspace*{2 cm}
\epsfig{file=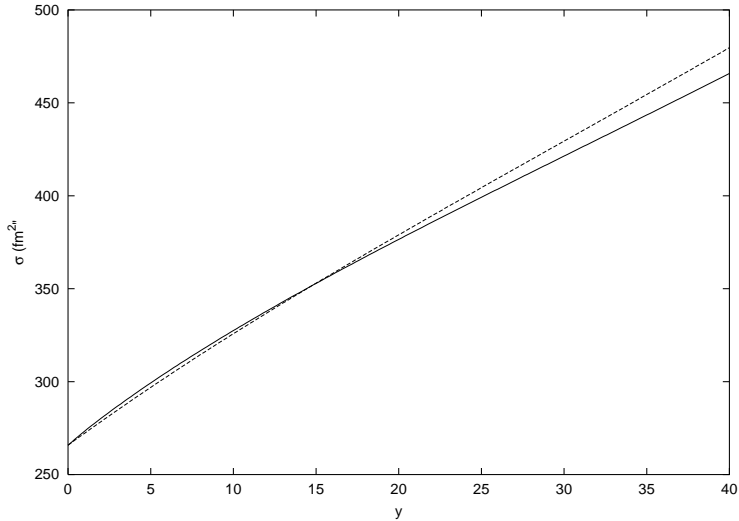,width=10 cm}
\caption{pPb cross-sections on the target nucleus with the Woods-Saxon
density and $\lambda=-0.069$ The lower (upper) curve on the right side
corresponds to presence (absence) of pomeron loops }
\label{fig6}
\end{figure}

The amplitude of oscillations is determined by the value of $\lambda$.
With smaller $\lambda$'s the oscillations die out. To see that we
repeated our calculations with a smaller value of the triple pomeron
coupling constant found in \cite{gotsman} $\lambda=-0.069$, that is
nearly ten times smaller than in (\ref{par}). With such $\lambda$
the perturbation parameter drops to $\zeta\sim 1/5$ inside the nucleus,
so that perturbative expansion becomes reasonable. Our results for
pPb cross-sections with $\lambda=-0.069$ are shown in Fig. \ref{fig6}.
We observe that the influence of loops is quite small. It grows with
rapidity and becomes noticeable at $y\sim 30\div 40$ only to vanish
in the end at still higher rapidities. Comparing with
Figs. \ref{fig4} and \ref{fig5} one also observes that with smaller
$\lambda$ the cross-sections rise faster with $y$ due to the behaviour
of the zero-order term (\ref{amp0}). At very small $\lambda$ the
damping exersized by fanning is small and ${\cal A}^{(0)}$ grows like
the pomeron itself.

\section{Conclusions}
We studied the contribution of the two simplest loops in the local
reggeon field theory with a supercritical pomeron in the nucleus.
We applied a quisi-local approximation which assumes that the slope
$\alpha'$ is small and can be put to zero in all places except in the
loops where the leading term at $\alpha'\to 0$ has been retained.
Our results confirm the conclusion reached in \cite{BT} that
the nuclear surrounding  transforms
the supercritical pomeron with the intercept $\alpha(0)-1=\epsilon>0$
into the subcritical one with the intercept $-\epsilon$. As a result,
at high energies the pomeron Green function vanishes and contributions
from multipomeron exchanges vanish still faster according to the standard
predictions for the subcritical pomeron. However this phenomenon becomes
effective at different energies depending on the nuclear density.
At small densities the Green function begins to decrease at very high
rapidities far beyond our present experimental possibilities. And in the
intermediate region loop contributions may become large if the triple
pomeron coupling constant $\lambda$ is not small enough.

So while the general conclusion about the validity of the perturbative
treatment is true, its practical realization depends on the value of
$\lambda$ and is determined by the perturbative parameter $\zeta$ defined
in (\ref{zeta}). Unfortunately with the standard values (\ref{par})
for $\lambda$ parameter $\zeta\sim 1$ and perturbative treatment is
dubious. Higher order loops have to be included into the picture to
have quantitative results. With smaller values of $\lambda$ the
loop contribution becomes small and perturbation approach is justified.

The influence of loops shows itself in oscillations in rapidity
of the cross-sections. These oscillations are due to the complex character
of the pomeron pole once loops are included. Due to the wrong sign of the
pomeron self-mass the pomeron poles does not move to the second sheet of the
energy plane as with normal particles but stays on the physical sheet
acquiring a nonzero imaginary part. Of course with very small $\lambda$
oscillations are small and have a very long pariod in $y$. However with
greater $\lambda$ they are clearly visible, as illustrated in Figs.
\ref{fig4} and \ref{fig5}

On the practical side our results point to yet another possibility
to experimentally measure the value of the
triple-pomeron coupling constant $\lambda$ on which the behaviour of the
cross-sections with energy critically depends. To exclude the region of
impact parameters $b$ well outside the nucleus, where our treatment
becomes invalid, one may select events in which the collisions occur
inside the nucleus by requiring the multiplicity of produced particles
to be large enough.

From the theoretical point of view we consider our results to be
promising  as a basis
for treating loop contributions in the perturbative QCD. It seems to
be advantageous to study them in the nuclear surrounding, which makes
the perturbative approach at high energies  much more tractable.

\section{Acknowledgments}
This work has been supported by grants RFFI 09-012-01327-a and RFFI-CERN
08-02-91004.

\end{document}